\documentstyle[12pt,draft]{article}
\parskip =\medskipamount
\newfont{\ensmathquatorze}{msbm10 scaled 1400}
\newfont{\ensmathonze}{msbm10 scaled 1100}
\newfont{\ensmathdix}{msbm10}
\newfont{\ensmathneuf}{msbm10 scaled 833}
\newfont{\ensmathhuit}{msbm10 scaled 694}
\newfam\ensmathfam                        
\textfont\ensmathfam=\ensmathonze        
\scriptfont\ensmathfam=\ensmathdix       
\scriptscriptfont\ensmathfam=\ensmathhuit

\def\be{\begin{equation}}
\def\ee{\end{equation}}
\def\bea{\begin{eqnarray}}
\def\eea{\end{eqnarray}}

\newcommand\li{\lambda_i}
\newcommand\lj{\lambda_j}
\newcommand\vli{\vert\lambda_i\vert}
\newcommand\vlj{\vert\lambda_j\vert}
\newcommand\tdz{^t\partial_z}
\newcommand\dzi{\partial_{z_i}}
\newcommand\dzib{\partial_{\bar z_i}}
\newcommand\dz{\partial_z}
\newcommand\dzb{\partial_{\bar z}}
\newcommand\tdzb{^t\partial_{\bar z}}
\newcommand\ti{\theta_i}
\newcommand\tj{\theta_j}
\newcommand\bb{{\bf B}}

\newcommand\gM{{\bf M}}
\newcommand\gR{{\bf R}}
\newcommand\mj{\mu_j}
\newcommand\mpj{\mu'_j}
\newcommand\ddz{{\cal D}z}
\newcommand\ddzb{{\cal D}{\bar z}}
\newcommand\oi{\omega_i}
\newcommand\oj{\omega_j}
\newcommand\ai{\alpha_j}
\newcommand\bm{\beta_{jm}}
\newcommand\gr{\gamma_{jr}}

\begin{document}


\title{  \bf  Statistical properties of the 2D attached Rouse chain}

\author{ Olivier B\'enichou$^{\dagger ,\star }$ and Jean Desbois$^{\dagger }$ }

\maketitle	

{\small
\noindent$^\dagger $ 
Laboratoire de Physique Th\'eorique et Mod\`eles Statistiques.
Universit\'e Paris-Sud, B\^at. 100, F-91405 Orsay Cedex, France.
}

{\small
\noindent$^\star$ 
Laboratoire de Physique Th\'eorique des Liquides, UPMC, 4 place Jussieu
 75252 Paris Cedex 05 
}
\vskip1.3cm

\begin{abstract}
 
We study various dynamical properties (winding angles, areas) of a set of
harmonically bound Brownian particles (monomers), one endpoint of this
chain being kept fixed at the origin 0. In particular, we show that, for
long times $t$, the areas $\{A_i\}$
enclosed by the monomers scale like $t^{1/2}$,
with correlated gaussian distributions. This is at variance with the winding
angles $\{\theta_i\}$ around fixed points that scale like $t$ and are
distributed according to independent Cauchy laws. 
 
 \end{abstract}
\vskip1.3cm

In this paper, we will study the planar motion of a chain of $n$ harmonically
bound brownian particles. This model is usually refered in the
litterature as the Rouse chain \cite{1} and has shown
to be historically very important in polymer science 
\cite{2,3}. We will consider such a chain attached at the origin 0 and examine
some of its properties from the Brownian motion viewpoint. 
\vskip.2cm
Representing a given configuration of the chain by a complex $n$-vector
$z$ (the components $ z_i,$ $i=1,\dots  ,n$, are
the complex coordinates of the particles), we consider the set of all the
closed trajectories of
length $t$, i.e. $ z(t)=z(0)$, and this for all the
starting configurations $z(0)$. Practically, we will not weight the
starting configurations with any thermodynamical factor. We are aware that
this approach is quite different from the one taken in polymer physics 
\cite{4} where, at $t=0$, the chain is supposed to
be in equilibrium with the environment at some finite temperature $T$.
\vskip.4cm
$A_j$ and $\theta_j$ being the area enclosed by
the $j^{th}$ particle and its winding angle around 0, our goal
is to compute the joint probability distributions $P(
\{A_i\})$ and $P(\{
\theta_i\})$ for such
trajectories. In order to make comparisons, we now recall some of the
results concerning the planar Brownian motion.
\vskip.3cm
We first quote the area and winding angle distributions, respectively
$P(A)$ (L\'{e}vy's law \cite{5}) and $P(
\theta)$
(Spitzer's law \cite{6}) for a particle allowed to
wander everywhere in the plane:
  
\be\label{1}  
   P(A)=\frac{\pi}{2t} \, \frac{1}{\cosh^2\frac{\pi A}{t}}
\ee   

\be\label{2}
   P(\theta )=\frac {2}{\pi\ln t}  \, 
   \frac {1}{1+ \left(\frac {2\theta}{\ln t} \right)^2 } 
\ee
(the last one holds, in the limit $t\to\infty$, 
for open curves, the final point being left unspecified). 
\vskip.2cm
Those two laws were obtained more 
than 40 years ago and since that 
time many refinements have been brought.
For instance, in \cite{7}, the authors pointed 
out the importance of the small windings occuring when the 
particle is close to 0. Excluding 
an arbitrary small zone around 0, they showed that the variance
$\langle\theta^2\rangle$ becomes 
finite in contrast with the Spitzer's result, eq.(\ref{2}).
\vskip.2cm
On the other hand, for Brownian motion on bounded
 domains \cite{8,9}, the scaling variables in the 
limit $t\to\infty$, become, resp., $A/\sqrt{t}$ and $\theta/t$
with still an infinite variance $\langle\theta^2\rangle$.
 We close here this brief recall and start our chain study 
 with the following set 
 of coupled Langevin equations :

\bea
      \dot z_1 &=& k \, (z_2 -2z_1) +\eta_1 \nonumber \\
  \dot z_l   &=& k \, (z_{l+1} + z_{l-1} -2z_l ) +\eta_l \ , \quad 
  2\le l \le n-1  \label{3} \\
  \dot z_n &=&   k \, (z_{n-1}-z_n) +\eta_n \nonumber  
\eea
where $k$ is the spring constant and
$\eta_m$ ($\equiv\eta_{mx}+i\eta_{my}$) a gaussian white noise:

\bea
	\langle \, \eta_m(t) \,	\rangle   &=& 0  \nonumber \\
   \langle \, \eta_m(t)\, \eta_{m'}(t') \, \rangle
 &=& 2 \, \delta_{mm'} \, \delta(t-t')     \label{4}
\eea

Introducing the complex $n$-vector
 $\eta $, eq. (\ref{3}) can be written in a matrix form:

\be\label{5}
        \dot z=- \;  k\; \gM\; z+\eta
\ee
where $\gM$ is the tridiagonal $(n\times n)$ matrix:

$$
\gM = \left(
\begin{array}{ccccc}
 \ 2 & -1  & \ 0  & \cdots & \ 0  \\
 -1  & \ 2  & -1  &  \cdots  & \ 0  \\
  \ 0  & -1  & \ 2  &  \cdots  & \ 0   \\
  \vdots & \vdots   & \vdots  & \ddots & \vdots  \\
 \ 0  & \ 0  & \ 0  & \cdots & \ 1  
\end{array}
\right)
$$
with an inverse given by:

$$
\gM^{-1} = \left(
\begin{array}{ccccc}
 1 & 1  & 1  & \cdots & 1  \\
 1  & 2  & 2  &  \cdots  & 2  \\
  1  & 2  & 3  &  \cdots  & 3  \\
  \vdots & \vdots   & \vdots  & \ddots & \vdots  \\
 1  & 2  & 3  & \cdots & n  
\end{array}
\right)
$$

The eigenvalues of $ \gM$ are:

\be\label{8}
   \oj \, = \, 2 \, \left( 1-\cos\frac{\pi (2j-1)}{2n+1}  \right)
 \   ,  \quad  1 \le j  \le n   
\ee
With the matrix $ {\bf \omega} ={\rm diag}(\omega_i)$, we can write:

\bea\label{9}
          {\bf \omega } &=& \gR^{-1}\; \gM\; \gR \\ 
\label{10} z &=& \gR \ Z    
\eea
where $\gR$ is an orthogonal matrix 
and the components of $Z$ are the normal coordinates. 

Let us call ${\cal P}( z, z_0,t)$ the 
probability for the chain to go from
$z_0$ at $t=0$ to $z$ at time $t$.
${\cal P}$ satisfies a Fokker-Planck equation \cite{10}:

\be\label{11}
   \partial_t {\cal P}=\left( \tdz \, k\, \gM\, z +
   \tdzb\, k\, \gM\, \bar z +2 \, \tdzb \, \dz
\right) {\cal P}
\ee
where $\dz$ (resp. $\dzb$) is a $n$-vector of components $\dzi$
 (resp. $\dzib$) and $\tdz$ (resp. $\tdzb$) is the transpose of
 $\dz$ (resp. $\dzb$). The solution can be written in terms of
a path integral ($\ddz \, \ddzb \, = \prod_{i=1}^n
{\cal D}{z_i}  {\cal D}{\bar {z_i}}$):

\bea
  {\cal P}(z,z_0,t) &=& \det \, \left( \, e^{tk\gM } \, \right) 
  \, \int_{z(0)=z_0}^{z(t)=z}\ddz\ddzb 
\exp\left( -\frac{1}{2}\int_0^t \,  ^t(\dot {\bar z} +k\gM \bar z)(\dot z +
 k\gM z)
  \, d\tau \right)
 \label{12}       \\
&\equiv & F(z,z_0,t).G(z,z_0,t) \nonumber  
\eea

with

\bea 
           F(z,z_0,t)    &=&  \det \, \left( e^{tk\gM } \right) \, 
	   e^{-\frac{1}{2}
     \left( ^t\bar z k\gM z   -  ^t\bar z_0 k\gM z_0         \right)     }     
\nonumber	    \\ 
  G(z,z_0,t)  &=&  \int_{z(0)=z_0}^{z(t)=z}\ddz\ddzb 
\exp\left( -\frac{1}{2}\int_0^t \left( ^t\dot {\bar z}\; \dot z \, +\, k^2
\;  ^t\bar z\;  \gM^2 z\right) d\tau \right)= 
\left\langle z\vert e^{-tH_0} \vert z_0
 \right\rangle      \\
       &=&  \det\left( \frac{{\bf S}}{2\pi } \right)
         \exp \left( -\frac{1}{2}
     \left(\,  ^t{\bar z}\, {\bf C}\, z +\, ^t{\bar z_0}\, {\bf C}\, z_0  
 \, -^t{\bar z}\, {\bf S}\, z_0 - \, ^t{\bar z_0}\, {\bf S}\, z \right) \right)
 \label{13} \\
      H_0   &=& -2 \, \tdzb\; \dz +\frac{1}{2} \, k^2 \ ^t{\bar z}
      \,  \gM^2\, z   \label{H0} 
\eea 

The matrices $\bf S$  and $\bf C$ appearing in (\ref{13}) are  defined as: 

\be\label{SC}
{\bf S} =  k\,\gM \; \left( \sinh (\, t\, k \; \gM\, )\right)^{-1}  \quad , 
\quad {\bf C} =  k\, \gM\; \coth (\, t\, k \; \gM\, )  
\ee

In fact, ${\cal P}$, eq. (\ref{12}),
can be easily deduced from the gaussian
distribution of $\eta$ (use  (\ref{5});
det($e^{tk \gM }$) is simply the functional
Jacobian for the change of variable $\eta\to z$ \cite{11}).

(\ref{13}) is a generalization of the
harmonic oscillator propagator \cite{12}.
 It is obtained by using the normal coordinates. Furthermore, as can be
 easily checked, ${\cal P}$ is properly normalized:
 $\int {\rm d}z{\rm d}{\bar z}\;{\cal P}(z,z_0,t)=1$.
\vskip.4cm
Remark that an effective measure can be built 
for a distinguished monomer of the chain 
\cite{4}: this can be done by integrating 
the Wiener measure (\ref{12}) over 
all the paths of the other monomers. 
The result is a complicated expression that contains, in particular, 
a non local part (in time) exhibiting the 
non-Markovian character of the process for this monomer. Nevertheless, 
we will show, in the sequel, that, 
despite this complication, we can compute 
some joint laws for several monomers (and {\it a fortiori}
for one monomer).
\vskip.7cm
So, let us turn to the computation of 
the area distribution $P(\{A_i\})$ 
for closed trajectories. Inserting the constraint

\be\label{14}
\prod_{j=1}^n \delta \left(A_j -\frac{1}{4i}
\int_0^t(z_j  \dot{\bar z_j} -  \bar z_j \dot {z_j}) d\tau \right)
\ee

in the Wiener measure and using 
$\delta(x)=\frac{1}{2\pi}\int e^{iBx}{\rm d}B$, 
we get the lagrangian for $n$ particles submitted to 
uniform magnetic fields orthogonal to the motion plane 
(in addition to the harmonic interactions). 
Remark that, in principle,
the magnetic fields are not the same for all the particles. 

Introducing the $(n\times n)$ diagonal 
matrix $\bf B$ (${\bf B}_{ij}=B_i\delta_{ij}$), we obtain

\bea
 P(\{ A_i \}) &=&  \int
 \left(  \prod_{j=1}^n \frac{dB_j}{2\pi } e^{iB_jA_j}
   \right)
   \left( \frac{ Z_{\bb }(t)}{Z_0(t)} \right) \label{15}  \\
       Z_{\bb }(t)   &=& \mbox{Tr} \,  e^{-tH_{\bb }}    \nonumber  \\
       H_{\bb }     &=&  H_0 + V            \nonumber  \\
    V  &=& \frac{1}{2}\left( - ^tz\, \bb\, \dz +  ^t\bar z\, \bb \, \dzb
    \right)   +\frac {1}{8} \,   ^t{\bar z} \, \bb^2 \, z    \label{16}
\eea

In general, the matrices ${\bf B}$ and ${\bf M}$ 
do not commute and it is a difficult task 
to get the partition function $Z_{\bb }(t)$.
On the other hand, the distribution of 
the total area $A=\sum_{i=1}^n A_i$ 
is obtained by taking $B_j=B$ for all $j$. 
In this case, ${\bf B}$ and ${\bf M}$ 
commute. Using normal coordinates and 
known results about the partition 
function of the ``2D harmonic oscillator + uniform magnetic field'' problem
\cite{13}, we get the characteristic 
function of $A$ (${\bf I_n}$ is 
the $(n\times n)$ unit matrix):

\bea
 \frac{ Z_{\bb }(t)}{Z_0(t)} &=& \prod_{j=1}^n \left( 
 \frac{\cosh (tk\oj) -1} {\cosh (t\sqrt{(k\oj )^2+(\frac{B}{2})^2})-
 \cosh (t\frac{B}{2})}\right) \label{17} \\
 &=& \frac{\det\left(  \cosh (tk\gM) - {\bf I_n} \right)}
 {\det\left(\cosh ( t\sqrt{(k\gM )^2+(\frac{\bb }{2})^2  }) - 
 \cosh (t\frac{\bb }{2} )
\right)}
\eea
(the $\omega_j$'s are defined in (\ref{8})). By Fourier transformation, the
above $j^{th}$ factor gives:

\bea
 P_{\oj }(A)  &=& 4\frac{\ai }{\pi }\sinh^2\left(\frac {\ai t}{2}\right) 
 \sum_{m,r=0}^{\infty } \left( \sqrt{\frac{\gr }{\bm }}
 K_1(2\sqrt{\bm\gr }) +  
 \sqrt{\frac{\bm }{\gr }}  K_{-1}(2\sqrt{\bm\gr })  \right) 
 \label{18} \\
 \gr  &=& \ai t (r+ 1/2) + i \, \ai A        \nonumber   \\
 \bm  &=& \ai t (m+ 1/2) - i \, \ai A        \nonumber   \\
 \ai  &=& k \, \oj    \nonumber   
\eea
where the $K_{\pm1}$ are modified Bessel functions. 

Thus, in the general case, $P(A)$ can be obtained by a $n$-convolution
product of the $P_{\omega_j}$'s. However, we are afraid that the final result 
 could be awkward! 
Nevertheless, 
if we consider the limit $t\to\infty$,
(\ref{17}) leads to:

\be\label{19}
 \frac{ Z_{\bb }(t)}{Z_0(t)} \sim \exp \left( -\frac{tB^2}{8k}
 \, \sum_{i=1}^n\frac{1}{\oi } \, \right) = \exp \left( -\frac{tB^2n(n+1)}{16k}
\right)
\ee

Then, Fourier transformation shows that, in the large $t$ limit, $A$ is
gaussian and scales like $\sqrt{t}$. Such a scaling is expected for
all the areas $A_i$. This is what we will demonstrate by
perturbation theory. When $t\to\infty$, we have

\be\label{20}
      Z_{\bb }(t) \sim e^{-tE_0(\bb )}
\ee
where $E_0(\bb )$ is the ground state energy. Moreover, due to the large
oscillations of the factor $e^{iB_jA_j}$ in (\ref{15}) when $A_j\to\infty$,
only small values of $B_j$ will contribute. So, it is enough to
compute $E_0(\bb )$ at lowest order in $\bb $. We will use the normal
coordinates $Z_i$.

The eigenstates of $H_0$ are given by \cite{14}

\be\label{21}
     \Psi_{\{ m_j\} ,\{ n_j\} } (\{ Z_j \} )
     =  \prod_{j=1}^n \left( \sqrt{
     \frac{\oj \, n_j!}{\pi (n_j + \vert m_j \vert )!}} e^{i \, m_j\tj }
( \oj \vert Z_j \vert^2 )^{  \vert m_j \vert/2  } 
    L_{n_j}^{ \vert m_j \vert } (\oj \vert Z_j\vert^2 ) 
    e^{-\frac{1}{2}\oj \vert Z_j\vert^2}  
     \right)   
\ee
\be\label{EN}
    E_{\{ m_j\} ,\{ n_j\} }    
    =  \sum_{j=1}^n (2n_j+ \vert m_j \vert +1)\oj  
\ee

where $L_{n_j}^{\vert m_j\vert }$ is a Laguerre polynomial and the ground
state is $\Psi_{\{ 0\} ,\{ 0\} }$. The perturbation $V$,
(\ref{16}), writes:

\be\label{22}
   V = \frac{1}{2}\left( - \, ^tZ\gR^{-1}\bb \gR\partial_Z 
    +  ^t\bar Z \gR^{-1}\bb \gR\partial_{\bar Z}       \right) +
    \frac{1}{8} \, ^t\bar Z \gR^{-1}\bb^2 \gR Z
\ee

At first order in $V$, we get:

\be\label{23}
  \Delta E_0^{(1)}(\bb ) = \int 
  \Psi^*_{\{ 0\} ,\{ 0\} } \,  V \,  \Psi_{\{ 0\} ,\{ 0\} } =
  \frac{1}{8k} \mbox{Tr} \, (\bb^2 \gM^{-1})=  \frac{1}{8k}
  \sum_{m=1}^n m \, B_m^2
\ee

Quadratic terms in $\bb $ will also be produced at second order in
$V$. The non-vanishing contributions will come out from the
transitions from the ground-state to the states
$\{m_j=\pm1,\;m_l=\mp 1,\; m_i=0\;\mbox{if}\;i\neq j,l\},\;\{n_k=0\}$. The
computation gives:

\bea
   \Delta E_0^{(2)}(\bb ) &=& -\frac{1}{16k} \sum_{m,m'=1}^n B_m B_{m'}
   \sum_{j\ne l} R_{ml}  R_{mj}  R_{m'l}  R_{m'j}
   \left(\frac{\frac{\omega_l }{\omega_j} + \frac{\omega_j}{\omega_l}}
   {\omega_l+ \omega_j}   \right)   \label{24}  \\
   &=&  -\frac{1}{8k} \sum_{m=1}^n m \, B_m^2 + 
 \frac{1}{2k}  \sum_{m,m'=1}^n B_m D_{m,m'} B_{m'}    \label{25}   
\eea

with: 

\be\label{26}
 D_{m,m'}  =  \frac{1}{4} \int_0^{\infty } \left[  \left(  e^{-\tau \gM }
  \right)_{m,m'} \right]^2 d\tau + \frac{1}{8} \sum_{l=1}^n
  \frac{ R_{ml}^2  \, R_{m'l}^2}{\omega_l}
\ee      

So, to lowest order in $\bb $, we get:

\be\label{27}
 \frac{ Z_{\bb }(t)}{Z_0(t)} \sim \exp\left( -\frac{t}{2k}
 \sum_{m,m'=1}^n B_m D_{m,m'} B_{m'} \right)
\ee
As can be easily checked, (\ref{19}) is recovered if we
set $B_m=B,\;\forall m$.

With (\ref{15}), we arrive at the probability distribution:

\be\label{28}
  P(\{ A_i \} ) = \left( \frac{k}{2\pi t}  \right)^{n/2} \frac{1}
  {\sqrt{\det D }}
  \exp\left( -\frac{k}{2t}\sum_{m,m'=1}^n A_m(D^{-1})_{m,m'}A_{m'}\right)
\ee

Thus, we observe that the areas $A_i$ are correlated gaussian
variables and that they scale like $t^{1/2}$ as expected. For the
special case $n=2$, we have:

\be\label{29}
P(A_1,A_2) = \sqrt{\frac {5}{3}}\frac{2k}{\pi t}
\exp\left( - \frac{2k}{9t}
 \left( 23 A_1^2 - 14 A_1 A_2 + 8 A_2^2 \right) \right)
\ee
The width of $A_2$ is larger than the one of $A_1$: this is related to
the fact that the second  particle is, in average, farther from 0
than the first one. So, it sweeps larger areas.
\vskip1cm
Now, going to the winding angles $\{\theta_i\}$ around 0, we proceed as
before and insert the constraint

\be\label{30}
\prod_{j=1}^n \delta \left(\tj -\frac{1}{2i}
\int_0^t \left( \frac {z_j  \dot{\bar z_j} -  \bar z_j \dot {z_j}}  
  {z_j\bar z_j}    \right) d\tau \right)       
\ee
in the Wiener measure. We are now faced to the problem of $n$
harmonically bound particles submitted to the magnetic fields of
point-like vortices located at the origin. The corresponding
hamiltonian is:

\bea
  H_{\lambda }  &=& H_0 + W                             \\
        W  &=& \sum_{i=1}^n\li \left( \frac{1}{z_i}\dzib -  
	\frac{1}{\bar z_i}\dzi \right) + 
	\sum_{i=1}^n \frac{\li^2}{2z_i\bar z_i} 
	\label{31}
\eea

and the distribution $P(\{\theta_i\})$ is given by:

\be\label{32}
 P(\{ \ti \}) = \int\left(  \prod_{j=1}^n \frac{d\lj }{2\pi }
 e^{i\lj\tj } \right)
   \left( \frac{ Z_{\lambda }(t)}{Z_0(t)} \right)
\ee

Studying  the limit $t\to\infty$, we cannot develop directly as before a
perturbation theory with $W$: this is because of the last term in $W$
that leads to a singular perturbation \cite{9}. Due to this term, all
the eigenfunctions of $H_\lambda$ must vanish in 0 at least as
$\prod_{i=1}^n(z_i {\bar z_i})^{\vli /2}\;(\equiv U)$
. So we redefine those
eigenfunctions \cite{9}:

\be\label{33}
       \Psi = U \widetilde{\Psi }
\ee

The new hamiltonian acting on $\widetilde{\Psi}$ is:

\bea
  \widetilde{H}  &=& H_0 +    \widetilde{W}   \label{34} \\ 
 \widetilde{W} &=&  \sum_{i=1}^n
 \left(  (\li - \vli ) \frac{1}{z_i}\dzib -(\li +\vli ) 
 \frac{1}{\bar z_i}\dzi
 \right) \label{35} 
\eea

That time, we can compute $\Delta E_0(\lambda)$ perturbatively and it will
appear that only first order is necessary. Integrals of the form

\be\label{36}
\int e^{-\frac{1}{2} \, ^t\bar z k\gM z}  \frac{1}{{\bar z_i}}\dzi  
   e^{-\frac{1}{2} \, ^t\bar z k\gM z} dz d\bar z
\ee
are involved. Integrating by parts and using 
 $\dzi\left(\frac{1}{\overline{z}_i}\right)=\pi\delta\left(z_i\right)$,
 we get, after some algebra:

\be\label{37}
\Delta E_0(\lambda )= k  \sum_{j=1}^n \frac{\vlj }{(\gM^{-1})_{jj}} =
k \sum_{j=1}^n  \frac{\vlj }{j} \equiv  \sum_{j=1}^n \mj\vlj
\ee

So, for the winding angle distribution, we obtain:

\bea
      P(\{ \ti \}) &=& \int\left(  \prod_{j=1}^n \frac{d\lj }{2\pi }
 e^{i\lj\tj } \right)   e^{-t  \sum_{j=1}^n \mj\vlj }   \label{38} \\
               &=&   \prod_{j=1}^n\left( \frac{1}{\pi\mj t}  
	     \,  \frac{1}{1+ (\frac{\tj }{\mj t})^2} \right)	  \label{39}
\eea

At large times, the winding angles are uncorrelated, they scale like
$t$ and are distributed according to Cauchy laws. The variance
$\langle\theta_j^2\rangle$ is infinite: this is, of course, due to the ``small
windings'' occuring in the vicinity of the origin as will be seen
explicitly at the end of this paper.

Moreover, we observe that $\theta_j$
scales like $\mu_j$, i.e. like $1/j$. This is reasonable because, when
$j$ increases, the considered particle is, in average, farther from 0
and, consequently, its winding angle must decrease. What is somewhat
unexpected is such a simple dependence of $\theta_j$ on $j$.
\vskip.3cm
We also addressed the problem of winding angles around
$n$ different points of complex coordinates $b_l$, $l=1,\ldots n$.
\vskip.1cm
$\theta_j'$ being the angle wound by the particle $j$ around the point
 $b_j$, we obtained for the set of variables 
 $\{ \theta_j' \} $ the same joint law as (\ref{39})
 except for the change of $\mu_j$ into $\mu_j'$:

\be\label{40}
     \mpj = \mj e^{-\mj \vert b_j \vert^2 }
\ee

Owing to the rotationnal symmetry breaking when $b_j\neq0$, the winding
angles $\theta_j'$ are statistically reduced by the factor
$e^{-\mu_j\vert b_j\vert^2}$. 
Nevertheless, even for large $\vert b_j\vert$'s,
the variance $\langle(\theta'_j)^2\rangle$ is infinite.
\vskip.2cm
Setting all the $b_j$'s to 
zero, we recover (\ref{39}). This is what we will consider now and assume
that we count the winding angles $\theta_j$ only when $\vert z_j\vert>r_0$
(i.e. the so-called ``big windings'' \cite{7}). Still when
$t\to\infty$, the perturbation $W$, eq.(\ref{31}), can now be
used because $\lambda_j=0$ when $\vert  z_j\vert<r_0$. At first
order in $W$, the linear contributions in $\lambda_j$ will cancel. In the
limit of a small, but finite $r_0$, we get, for the remaining
contribution:

\be\label{41}
\Delta E_0^{(1)} (\lambda ) \sim k \vert  \ln r_0    \vert
  \sum_{j=1}^n \frac {\lj^2}{j}
\ee

The quadratic contributions  in the $\lambda_j$'s coming out from the
second order in $W$ will be finite (thus subleading) when
$r_0\to0^+$. Finally, we get for the  
big winding angles  asymptotic    distribution:

\be\label{42}
     P(\{ \tj \}) =  \prod_{j=1}^n \sqrt{\frac{j}{4\pi t 
    k \vert  \ln r_0 \vert } } \, \exp \left( -\frac{j }
    {4 t k \vert  \ln r_0 \vert } \tj^2 \right)
\ee

In this limit,the variables $\theta_j$ are uncorrelated: the correlations
get smaller and smaller when $r_0$ decreases. They are now gaussian
and scale like $\sqrt{t\vert \ln{r_0}\vert /j}$. Their variance
grows to infinity when $r_0$ goes to 0, showing the increasing
contribution of the small windings around 0.

\vskip.7cm

To summarize, we have computed explicitly the asymptotic joint laws for the
areas (that scale like $\sqrt{t}$) and for the winding angles (that
scale like $t$ when no critical region is excluded).
 The scaling variables  we have got 
compare well with those involved in  the Brownian motion on finite domains:
this  is not so surprising since  the chain is  bound to a fixed point.

Moreover, we have shown that physical interactions
between particles (harmonic interactions here) can lead to
statistical correlations (case of the areas) or not (case of the winding 
angles): it depends on the quantity  we consider. 

In a forthcoming paper \cite{15}, we will study the statistical properties
 of the free Rouse chain. We will especially show that the areas and winding
angles distributions are very different from those presented in this
work. This is essentially due to the translation invariance that holds
when the chain is free. 

\vskip1cm
One of us (O. B.) acknowledges  Dr. G. Oshanin  for drawing his attention to
this problem.

\vskip1cm
{\bf e-mail:}

benichou@lptl.jussieu.fr

desbois@ipno.in2p3.fr

\end{document}